\documentclass{elsart}                                               
\usepackage{amssymb}
\journal{Nuclear Physics B}
\begin{document}
\begin{frontmatter}
\title{Quantum supersymmetric Toda-mKdV hierarchies}
\author{Petr P. Kulish}
\ead{kulish@pdmi.ras.ru}
\author{Anton M. Zeitlin\corauthref{cor1}}
\ead{zam@math.ipme.ru} 
\ead[url]{http://www.ipme.ru/zam.html}
\address{St. Petersburg Department of Steklov Mathematical 
Institute, Fontanka, 27,\\ St. Petersburg, 191023, Russia}
\corauth[cor1]{Corresponding author}
\date{}
\begin{abstract}
In this paper we generalize the quantization procedure of 
Toda-mKdV hierarchies to the case of arbitrary affine (super)algebras. 
The quantum analogue of the monodromy matrix, related to the 
universal R-matrix with the lower Borel subalgebra 
represented by the corresponding vertex operators is introduced. 
The auxiliary L-operators satisfying RTT-relation are constructed 
and the quantum integrability condition is obtained. 
General approach is illustrated by means of two important examples. 
\end{abstract}

\end{frontmatter}
\section{Introduction}
\hspace*{5mm}During more than quarter of a century 
both classical and quantum 
affine Toda field theories and related generalized (m)KdV ((modified) Korteweg-de Vries) 
hierarchies were extensively studied (see e.g. \cite{DS}-\cite{frenkel}). 
These theories are integrable and have the L-A pair (or zero curvature) 
formulation. The most famous are sine-Gordon
and $A_2^{(2)}$ models (see e.g. \cite{solitons}, \cite{izkorbog}) 
and the associated 
KdV and reduced Boussinesq hierarchies. These theories allow the 
supersymmetric and fermionic generalizations \cite{olsh}-\cite{evans} 
both called ``super'' because their underlying algebraic 
structures are affine Lie superalgebras.\\ 
\hspace*{5mm}The super-Toda field theory appears to be 
supersymmetric if and only if the associated superalgebra possesses 
the purely fermionic system of simple roots. 
One should note, however, that usually the 
superalgebras allow a few simple root systems \cite{kac}, they correspond to 
different Toda theories and only purely fermionic system corresponds to the 
supersymmetric one.\\ 
\hspace*{5mm}The supersymmetric version of the Drinfeld-Sokolov reduction 
applied to the matrix L-operator of the Toda-mKdV theories gives the 
generators of the related super W-algebra with the commutation relations 
provided by the associated Hamiltonian structure (see e.g. 
\cite{gomes}, \cite{delduc}).\\  
\hspace*{5mm}In this paper we consider these theories from a point of view of 
quantum inverse scattering method (QISM) 
(see \cite{leshouches}, \cite{kulsklyan}). 
There are two approaches of applying QISM to Toda-mKdV type models.
The first one is more traditional and based on the quantization of the corresponding 
lattice systems (see e.g. \cite{fv},\cite{volkov}) and 
the second one is based on the quantization in terms of continuous free 
field theory and was introduced in \cite{lukyanov}, \cite{1-3}.\\ 
\hspace*{5mm}Here we use the second approach, generalizing our 
results obtained in \cite{plb1}-\cite{npb} for the affine superalgebras 
of rank 2 to the case of general affine superalgebra.\\ 
\hspace*{5mm}We build the quantum generalization of the 
monodromy matrix and prove that the related auxiliary $\mathbf{L}$-operators (which are equal 
to the monodromy matrix multiplied by the exponential of the elements from the Cartan 
subalgebra) satisfy the RTT-relation \cite{leshouches}, \cite{kulsklyan}, while the 
quantum counterpart of the monodromy matrix itself is shown to satisfy the 
specialization of the reflection equation (see e.g. \cite{reflection}). 
This provides the quantum integrability relation for the supertraces of the 
monodromy matrix taken 
in different representations (transfer matrices). Moreover, it is proven that the 
auxiliary $\mathbf{L}$-operators are related with the universal R-matrix associated 
with the underlying quantum affine superalgebra, 
with the lower Borel subalgebra represented by the vertex operators from the corresponding Toda 
field theory. Using this relation in the case when the 
simple root system is purely fermionic, it is demonstrated 
that the transfer matrix is invariant under the 
supersymmetry transformation as it was shown on particular classical examples of Toda field 
theories \cite{penati}, \cite{evans}.\\ 
\hspace*{5mm}
In the last two sections the above constructions are illustrated by 
means of two important examples of integrable hierarchies: quantum 
super-KdV \cite{plb1} and SUSY N=1 KdV \cite{plb2}. 
These hierarchies generate two integrable structures of the superconformal 
field theory, the second one is invariant under the SUSY transformation 
while the first 
one is not.
\section{Bosonic Toda-mKdV hierarchies}
Each mKdV hierarchy and related Toda field 
theory associated with affine Lie algebra are generated by the following L-operator \cite{DS}:
\begin{eqnarray}\label{lop}
\mathcal{L}=\partial_{u} 
-\partial_{u}\phi^i(u)H^{i}-(\sum_{i=0}^{r}e_{\alpha_i}),
\end{eqnarray}
where $u$ lies on a cylinder of circumference $2\pi$, $\phi^i$ are the scalar fields with 
the Poisson brackets:
\begin{eqnarray}\label{poisson}
\{\partial_{u}\phi^i(u),\partial_{v}\phi^j(v)\}=-\delta^{ij}\delta'(u-v)
\end{eqnarray}
with quasiperiodic boundary condition:
\begin{eqnarray}\label{qperiodic}
\phi^i(u+2\pi)=\phi^i(u)+2\pi i p^i.
\end{eqnarray}
$e_{\alpha_i}$ are the Chevalley generators of the underlying affine Lie algebra
and $H^{i}$ $(i=1,..., r)$ form a basis in the Cartan subalgebra of the 
corresponding simple Lie algebra:
\begin{eqnarray}\label{boscommrel}
[H^i, e_{\alpha_k}]=\alpha_k^i e_{\alpha_k}, \quad[e_{\alpha_k}, e_{-\alpha_l}]=\delta_{kl}h_{\alpha_k},
\quad ad_{e_{\pm\alpha_k}}^{1-a_{kj}}e_{\pm\alpha_j}=0,
\end{eqnarray}
where $a_{kj}$ is a Cartan matrix and $h_{\alpha_k}\equiv(\alpha_k,H)=\alpha_k^i H^i$. 
In our case this algebra is 
considered in evaluation representations, when $e_{\alpha_0}=\lambda e_{-\theta}$
(this corresponds to the case of untwisted affine Lie algebra, the 
twisted case is more complicated), $\theta$ is the highest root of the related simple Lie algebra.  
The classical monodromy matrix 
for the linear problem 
associated with the L-operator (\ref{lop}) 
can be expressed in the following way \cite{lukyanov}:
\begin{eqnarray}\label{bosmonmatr}
\pi_s(\lambda)(\mathbf{M})\equiv \mathbf{M_s(\lambda)}=
e^{2\pi i p^k H^k}Pexp\int_0^{2\pi}\d u(\sum^r_{i=0} e^{-(\alpha_i, \phi(u))}
e_{\alpha_i}),
\end{eqnarray}
where $\pi_s$ is some evaluation representation of the corresponding affine 
Lie algebra. Defining the auxiliary $\mathbf{L}$-matrix:
\begin{eqnarray}\label{LM}
\mathbf{L(\lambda)}=e^{-\pi i p^k H^k}\mathbf{M(\lambda)}
\end{eqnarray}
one can find that the quadratic Poisson bracket relation is valid:
\begin{eqnarray}\label{rpoisson}
\{\mathbf{L(\lambda)}\otimes, \mathbf{L(\mu)}\}=[\mathbf{r}(\lambda\mu^{-1}),
\mathbf{L(\lambda)}\otimes\mathbf{L(\mu)}],
\end{eqnarray}
where $\mathbf{r}(\lambda)$ is the trigonometric r-matrix \cite{kulsklyan}, 
\cite{BD} related with the corresponding simple Lie algebra. The traces of the 
monodromy matrices in different evaluation representations $\pi_s$:
\begin{eqnarray}\label{cltransfer}
\mathbf{t_s(\lambda)}=\pi_s (\mathbf{M(\lambda)})
\end{eqnarray}
are in involution under the Poisson brackets:
\begin{eqnarray}\label{clintegr}
\{\mathbf{t_s(\lambda)}, \mathbf{t_{s'}(\mu)}\}=0.
\end{eqnarray}
The quantization means that we move from the quadratic Poisson bracket 
relation (\ref {rpoisson}) to the RTT-relation with the underlying affine Lie 
algebra deformed to the quantum affine algebra (see e.g. \cite{leshouches}, \cite{kulsklyan} 
and below).
In this section we give the generalization of the constructions appeared in \cite{1-3}, \cite{5}.\\
\hspace*{5mm}
First, let's quantize the scalar fields $\phi^i$:
\begin{eqnarray}\label{freefields} 
&&\phi^k(u)=iQ^k +iP^k u+\sum_n\frac{a^k_{-n}}{n}e^{inu},\\
&&[Q^k,P^j]=\frac{i}{2}\beta^2\delta^{kj} ,\quad 
[a^k_n,a^j_m]=\frac{\beta^2}{2}n\delta^{kj}\delta_{n+m,0},\nonumber
\end{eqnarray}
and define the vertex operators
\begin{eqnarray}
&&\tilde{V}_{\alpha_k}(u):=:e^{-(\alpha_k,\phi(u))}:=
\exp\Big(\sum_{n=1}^{\infty}\frac{(-\alpha_k, a_{-n})}{n}e^{inu}\Big)
\nonumber\\
&&\exp\Big(-i((\alpha_k,Q)+(\alpha_k,P)u)\Big)
\exp\Big(\sum_{n=1}^{\infty}\frac{(\alpha_k,a_{n})}{n}e^{-inu}
\Big).\nonumber
\end{eqnarray} 
Then one can define the quantum generalizations of auxiliary $\mathbf{L}$-operators \cite{lukyanov}:
\begin{eqnarray}\label{qlop}
\mathbf{L}^{(q)}(\lambda)=
e^{\pi i P^k H^k}Pexp\int_0^{2\pi}\d u(\sum^r_{i=0} :e^{-(\alpha_i, \phi(u))}:
e_{\alpha_i}),
\end{eqnarray} 
where now $e_{\alpha_i}$, $H^k$ are the generators of the corresponding 
quantum affine algebra:
\begin{eqnarray}\label{commrel}
[H^i, e_{\alpha_k}]=\alpha_k^i e_{\alpha_k},\quad 
[e_{\alpha_k}, e_{-\alpha_l}]=\delta_{kl}[h_{\alpha_k}]_q,\quad 
ad_{e_{\pm\alpha_k}}^{{(q)}^{1-a_{kj}}}e_{\pm\alpha_j}=0,
\end{eqnarray}
where $q=e^{i\pi\frac{\beta^2}{2}}$, $[a]_q=(q^a-q^{-a})/(q-q^{-1})$ and 
$ad^{(q)}_{e_{\alpha}}e_{\beta}=e_{\alpha}e_{\beta}-
q^{(\alpha,\beta)}e_{\beta}e_{\alpha}$.
The object (\ref {qlop}) is defined in the interval 
$0<\beta^2<2b^{-1}$, $b=max(|b_{ij}|)$ $(i\ne j)$, where $b_{ij}$ is a symmetrized Cartan matrix, 
but can be analytically continuated to a wider region. The quantum monodromy matrix is 
defined using the relation (\ref{LM}): 
\begin{eqnarray}\label{qLM}
\mathbf{M}^{(q)}(\lambda)=e^{\pi i P^k H^k}\mathbf{L}^{(q)}(\lambda).
\end{eqnarray} 
It can be shown that the $\mathbf{L}^{(q)}$ operator satisfies the 
mentioned RTT-relation in the two ways \cite{1-3}, \cite{5}. 
The first way is to consider the following product: 
\begin{eqnarray}\label{LIIL}
(\mathbf{L}^{(q)}(\lambda)\otimes I)(I \otimes\mathbf{L}^{(q)}(\mu)).
\end{eqnarray}
Then, moving all the Cartan multipliers to the left we find the 
following:
\begin{eqnarray}\label{PexpPexp}
&&e^{i\pi P^i \Delta(H^i)}Pexp\int_0^{2\pi}\d u \tilde{K}_1(u)Pexp\int_0^{2\pi}\d u K_2(u),\\
&&K_1(u)=\sum^{r}_{j=0}:e^{-(\alpha_j, \phi(u))}:e_{\alpha_j}\otimes q^{h_{\alpha_j}},\quad
K_2(u)=\sum^{r}_{j=0}1\otimes :e^{-(\alpha_j, \phi(u))}:e_{\alpha_j},\nonumber
\end{eqnarray}
where $\Delta(H^i)=H^i\otimes I+ I\otimes H^i$. 
The commutation relations between vertex operators on a circle:
\begin{eqnarray}\label{VV}
\tilde{V}_{\alpha_k}(u)\tilde{V}_{\alpha_j}(u')=q^{b_{kj}}\tilde{V}_{\alpha_j}(u')\tilde{V}_{\alpha_k}(u),
\quad u > u',
\end{eqnarray}
lead to
\begin{eqnarray}
[\tilde{K}_1(u),K_2(u')]=0, \quad u<u'.
\end{eqnarray}
Due to this property one can unify two P-exponents into the single one which is equal to 
\begin{eqnarray}
\pi_s(\lambda)\otimes\pi_{s'}(\mu)\Delta(\mathbf{L}^{(q)}),
\end{eqnarray} 
where $\pi_s$ and $\pi_{s'}$ are some evaluation representations, 
and the coproduct $\Delta$ of the quantum affine (super)algebra is 
defined by:
\begin{eqnarray}
&&\Delta(H^i)=H^i\otimes I + I\otimes H^i,
\quad
\Delta(e_{\alpha_j})=e_{\alpha_j}\otimes q^{h_{\alpha_j}}+1\otimes 
e_{\alpha_j},\\
&&\Delta(e_{-\alpha_j})=e_{-\alpha_j}
\otimes 1+q^{-h_{\alpha_j}} \otimes e_{-\alpha_j}.\nonumber
\end{eqnarray}
Next, considering opposite product of the $\mathbf{L}$-operators, 
one finds that it coincide with opposite coproduct of the 
$\mathbf{L}$-operators:
\begin{eqnarray}\label{ILLI}
(I \otimes\mathbf{L^{(q)}(\mu)})(\mathbf{L^{(q)}(\lambda)}\otimes I)=
\pi_s(\lambda)\otimes\pi_s(\mu)\Delta^{op}(\mathbf{L^{(q)}}),
\end{eqnarray}
where $\Delta^{op}=\tau \Delta$ and the map $\tau$ is defined as follows: 
$\tau(a\otimes b)= b\otimes a$. Using the property of the universal R-matrix, namely 
$\mathbf{R}\Delta=\Delta^{op}\mathbf{R}$ \cite{drinfeld}, we arrive to the RTT-relation:
\begin{eqnarray}\label{RTT}
&&\mathbf{R}(\lambda\mu^{-1})(\mathbf{L}^{(q)}(\lambda)\otimes I)(I \otimes\mathbf{L}^{(q)}(\mu))=\\
&&(I \otimes\mathbf{L}^{(q)}(\mu))(\mathbf{L}^{(q)}(\lambda)\otimes I)
\mathbf{R}(\lambda\mu^{-1}).\nonumber
\end{eqnarray}
Remembering the expression for the monodromy matrix (\ref{qLM}) one obtains 
that the RTT-relation is no longer valid for the monodromy matrices, 
however, it is easy to see that multiplying both RHS and LHS of (\ref{RTT}) 
by $e^{i\pi \Delta{H^k} P^k}$ one obtains:
\begin{eqnarray}\label{RMM}
\mathbf{R}_{12}(\lambda\mu^{-1})
\mathbf{\tilde{M}}_1^{(q)}(\lambda)\mathbf{M}^{(q)}_2(\mu)=
\mathbf{\tilde{M}}_2^{(q)}(\mu)\mathbf{M}_1^{(q)}(\lambda)
\mathbf{R}_{12}(\lambda\mu^{-1}),
\end{eqnarray}
where we have denoted $\mathbf{\tilde{M}}_1^{(q)}(\lambda)$ the monodromy 
matrix with $e_{\alpha_i}\otimes 1$ replaced by 
$e_{\alpha_i}\otimes q^{-h_{\alpha_i}}$ and 
$\mathbf{\tilde{M}}_2^{(q)}(\lambda)$ the monodromy matrix with 
$ 1\otimes e_{\alpha_i}$ replaced by 
$q^{-h_{\alpha_i}} \otimes e_{\alpha_i}$. 
Taking the trace, the above additional Cartan factors in 
$\mathbf{\tilde{M}}^{(q)}$ cancel and we obtain the quantum 
integrability condition for their traces (transfer-matrices):
\begin{eqnarray}\label{qintegr}
[\mathbf{t_s(\lambda)}, \mathbf{t_{s'}(\mu)}]=0.
\end{eqnarray} 
Another more universal way to obtain the RTT-relation is the correspondence between
the reduced universal R-matrix (see below) and the P-exponential form of the auxiliary 
$\mathbf{L}^{(q)}$ operator. That is, let's consider integrals of vertex operators: 
\begin{eqnarray}
V_{\alpha_k}(u_2,u_1)=\frac{1}{q-q^{-1}}\int_{u_1}^{u_2}\d u\tilde{V}_{\alpha_k}(u).    
\end{eqnarray}
Via the contour technique \cite{feigin} one can show that these objects 
satisfy the quantum Serre relations of the lower Borel subalgebra of the 
associated quantum affine algebra 
with simple roots $\alpha_k$. Using the structure of the reduced universal R-matrix 
(see e.g. \cite{tolstkhor} and Appendix) we can write $\bar{R}=K^{-1}\mathbf{R}=
\bar{R}(\bar{e}_{\alpha_i},
\bar{e}_{-\alpha_i})$, where
\begin{eqnarray}
\bar{e}_{\alpha_i}=e_{\alpha_i}\otimes 1,\quad 
\bar{e}_{-\alpha_i}=1\otimes e_{-\alpha_i},
\end{eqnarray}
$\mathbf{R}$ is a universal R-matrix and $K$ depends on the elements from Cartan 
subalgebra, because $\bar{R}$ is represented as a power series of these elements.
Then, following \cite{5} and using the fundamental feature of the universal R-matrix: 
\begin{eqnarray}
(I\otimes\Delta)\mathbf{R}=\mathbf{R}^{13}\mathbf{R}^{12},
\end{eqnarray}
one can show that the reduced R-matrix has the following property: 
\begin{equation}
\bar{R}(\bar{e}_{\alpha_i},e'_{-\alpha_i}+e''_{-\alpha_i})=
\bar{R}(\bar{e}_{\alpha_i},e'_{-\alpha_i})\bar{R}(\bar{e}_{\alpha_i},
e''_{-\alpha_i}),
\end{equation} 
where 
\begin{eqnarray}
e'_{-\alpha_i}&+&e''_{-\alpha_i}=(I\otimes\Delta)(1\otimes e_{-\alpha_i}),\\
e'_{-\alpha_i}&=&1\otimes q^{-h_{\alpha_i}}\otimes 
e_{-\alpha_i},\quad e''_{-\alpha_i}=1\otimes e_{-\alpha_i}\otimes 1,\quad 
\bar{e}_{\alpha_i}=e_{\alpha_i} \otimes 1\otimes 1.\nonumber 
\end{eqnarray}
Their commutation relations are
\begin{eqnarray}
e'_{-\alpha_i}\bar{e}_{\alpha_j}&=&\bar{e}_{\alpha_j}e'_{-\alpha_i},\quad
e''_{-\alpha_i}\bar{e}_{\alpha_j}=\bar{e}_{\alpha_j}e''_{-\alpha_i},\\
e'_{-\alpha_i}e''_{-\alpha_j}&=&q^{b_{ij}}e''_{-\alpha_j}e'_{-\alpha_i}.
\nonumber
\end{eqnarray} 
Now, denoting by $\mathbf{\bar{L}}^{(q)}(u_2, u_1)$ 
the reduced R-matrix with $e_{-\alpha_i}$ represented by 
$V_{\alpha_i}(u_2,u_1)$ and using the above property of $\bar{R}$ with 
$e'_{-\alpha_i}$ replaced by appropriate vertex operators we find: 
\begin{eqnarray}
\mathbf{\bar{L}}^{(q)}(u_3, u_1)=\mathbf{\bar{L}}^{(q)}(u_3, u_2)
\mathbf{\bar{L}}^{(q)}(u_2, u_1),\quad u_3 \ge u_2\ge u_1.
\end{eqnarray} 
Hence, $\mathbf{\bar{L}}^{(q)}$ has the property of P-exponent. 
When the interval $\delta=[u_2, u_1]$ is small enough one can show that 
\begin{eqnarray}
\mathbf{\bar{L}}^{(q)}(u_2, u_1)=1+\int_{u_1}^{u_2}\d u(\sum^r_{i=0} :e^{-(\alpha_i, \phi(u))}
:e_{\alpha_i})+O(\delta^2).
\end{eqnarray}
That is, we obtain that 
\begin{eqnarray}
\mathbf{\bar{L}}^{(q)}(u_2, u_1)=Pexp\int_{u_1}^{u_2}\d u(\sum^r_{i=0} :e^{-(\alpha_i, \phi(u))}
:e_{\alpha_i})
\end{eqnarray}
and $\mathbf{L}^{(q)}=e^{i\pi H^i P^i}\mathbf{\bar{L}}^{(q)}(2\pi, 0)$
satisfies RTT relation by construction.
\section{Quantum P-exponential and Toda-mKdV hierarchies based on 
superalgebras}
Now let's generalize the above results to the case when the underlying algebraic 
structures and integrable hierarchies are related to the affine Lie superalgebra. 
In the previous part we have moved from classical 
theory to the quantum one, here we will go in opposite direction, moving from the quantum version 
of the monodromy matrix and 
related auxiliary $\mathbf{L}$-operators, satisfying RTT-relation to their classical counterparts.\\  
\hspace*{5mm}First, let's introduce two types of vertex operators, bosonic and fermionic ones:
\begin{eqnarray}
W_{\alpha_i}^F(u)&\equiv&\int\d \theta :e^{-(\alpha_i,\Phi(u,\theta))}:=
\frac{i}{\sqrt{2}}(\alpha_i,\xi(u)):e^{-(\alpha_i,\phi(u))}:\\
W_{\alpha_i}^B(u)&\equiv&\int\d \theta \theta:e^{-(\alpha_i,\Phi(u,\theta))}:=:e^{-(\alpha_i,\phi(u))}:,
\end{eqnarray}
where $\Phi^k$ are the superfields: 
$\Phi^k(u,\theta)=\phi^k(u)-\frac{i}{\sqrt{2}}\theta\xi^k(u)$ and $\theta$ is a 
Grassmann variable. 
Their commutation relations on a circle are:
\begin{eqnarray}\label{commcircle}
W_{\alpha_i}^s(u)W_{\alpha_k}^{s'}(u')=(-1)^{p(s)p(s')}q^{b_{ik}}
W_{\alpha_k}^{s'}(u')W_{\alpha_i}^s(u), \quad u>u',
\end{eqnarray}
where $b_{kj}$ is the symmetrized Cartan matrix for the corresponding affine Lie superalgebra, 
$s$, $s'$ are $B$, $F$ and $p(F)=1, p(B)=0$ .\\
\hspace*{5mm}The mode expansion for the bosonic fields is the same as in (\ref{freefields}) and 
for fermionic fields $\xi^k(u)$ is the following:
\begin{eqnarray}
\xi^l(u)=i^{-1/2}\sum_n\xi^l_n e^{-inu},\quad\{\xi^k_n,\xi^l_m\}=\beta^2\delta^{kl}\delta_{n+m,0}.
\end{eqnarray}
These fermion fields may satisfy two 
boundary conditions periodic and 
antiperiodic $\xi^i(u+2\pi)=\pm\xi^i(u)$ corresponding to the two sectors of (S)CFT -- Ramond 
(R) and Neveu-Schwarz (NS) (the supersymmetry operator appears only when all fermions are in 
the R sector).\\   
\hspace*{5mm}It can be shown that the integrals of the introduced vertex operators as 
in the bosonic case satisfy the Serre and ``non Serre'' relations (see e.g. \cite {scheunert}- 
\cite{tolstkhor}) for the lower Borel subalgebra:
\begin{equation}
ad_{e_{-\alpha_k}}^{{(q)}^{1-a_{kj}}}e_{-\alpha_j}=0,\quad
[[e_{\pm\alpha_r},e_{\pm\alpha_s}]_q,[e_{\pm\alpha_r},e_{\pm\alpha_p}]_q]_q=0, 
\end{equation}
where the $ad^{(q)}$ operator is defined in the following way:
$ad^{(q)}_{e_{\alpha}}e_{\beta}=e_{\alpha} e_{\beta}-(-1)^{p(\alpha)p(\beta)}
q^{(\alpha,\beta)}e_{\beta}e_{\alpha}$, and $e_{\pm\alpha_r}$ is a so-called ``grey'' root 
(for more details see Appendix) of the quantum affine superalgebra with the corresponding 
bosonic and fermionic roots $\alpha_i$.\\
\hspace*{5mm}Substituting them with the appropriate multiplier 
$(q-q^{-1})^{-1}$ in the reduced R-matrix 
one can find (easily generalizing the results of section 2 to the case of superalgebra) that 
it satisfies the P-exponential multiplication property:
\begin{eqnarray}\label{q-Pexp}
\mathbf{\bar{L}}^{(q)}(u_2, u_1)=
Pexp^{(q)}\int_{u_1}^{u_2}\d u(\sum_{f} W^F_{\alpha_f}(u)e_{\alpha_f}+
\sum_{b}W^B_{\alpha_b}(u)e_{\alpha_b})\\
\mathbf{\bar{L}}^{(q)}(u_3, u_1)=\mathbf{\bar{L}}^{(q)}(u_3, u_2)
\mathbf{\bar{L}}^{(q)}(u_2, u_1),\quad u_3 \ge u_2\ge u_1,\nonumber
\end{eqnarray}  
where indices $f$ and $b$ imply that we are summing over fermionic and 
bosonic simple roots. The letter $q$ over the $Pexp$ means that the 
object introduced above in some cases (more precisely when a number of fermionic roots 
is more than one) cannot be written as P-exponential for any value of the deformation 
parameter due to the singularities in the operator products generated by the fermion 
fields $\xi^i$.       
Thus we call this object quantum P-exponential.\\
\hspace*{5mm}Defining then $\mathbf{L}^{(q)}\equiv 
e^{\pi i p^i H^i}\mathbf{\bar{L}}^{(q)}(2\pi,0)$ we find (similarly to 
the purely bosonic case) that it satisfies the RTT 
relation (\ref{RTT}) and defining the monodromy matrix $\mathbf{M}^{(q)}\equiv 
e^{\pi i p^i H^i}\mathbf{L}^{(q)}$ we again 
arrive to the property (\ref{RMM}) 
and obtain again the quantum integrability condition (\ref{qintegr}) for 
$\mathbf{t}^{(q)}=str\mathbf{M}^{(q)}$. 
We mention here that the relation (\ref{RMM}) can be rewritten in a more universal way, 
as a specialization of the reflection equation \cite{reflection}:
\begin{eqnarray}
\mathbf{\tilde{R}}_{12}(\lambda\mu^{-1})
\mathbf{M}_1^{(q)}(\lambda)F^{-1}_{12}\mathbf{M}^{(q)}_2(\mu)=
\mathbf{M}_2^{(q)}(\mu)F^{-1}_{12}\mathbf{M}_1^{(q)}(\lambda)\mathbf{R}_{12}(\lambda\mu^{-1}),
\end{eqnarray}
where $F=K^{-1}$ 
the Cartan's factor from the universal R-matrix (see Appendix), 
and $\mathbf{\tilde{R}}_{12}(\lambda\mu^{-1})=
F^{-1}_{12}\mathbf{R}_{12}(\lambda\mu^{-1})F_{12}$.\\ 
\hspace*{5mm}Now let's analyse the classical limit of the defined objects.
We will use the P-exponential property of $\mathbf{\bar{L}}^{(q)}(2\pi,0)$.
Let's decompose $\mathbf{\bar{L}}^{(q)}(2\pi,0)$ in the following way:
\begin{equation}
\mathbf{\bar{L}}^{(q)}(2\pi,0)=
\lim_{N\to\infty}\prod_{m=1}^{N}\mathbf{\bar{L}}^{(q)}(x_{m},x_{m-1}),
\end{equation} 
where we divided  the interval $[0,2\pi]$ into infinitesimal 
intervals $[x_m,x_{m+1}]$
with $x_{m+1}-x_m=\epsilon=2\pi/N$.
Let's find the terms that can give contribution of the first order
in $\epsilon$ in $\mathbf{\bar{L}}^{(q)}(x_{m},x_{m-1})$. 
In this analysis one needs the operator product expansion of fermion fields and 
vertex operators:
\begin{eqnarray}
&&\xi^k(u)\xi^l(u')=
-\frac{i\beta^2\delta^{kl}}{(iu-iu')}+\sum_{p=0}^{\infty}c^{kl}_p(u)(iu-iu')^p,\\
&&:e^{-(\alpha_k,\phi(u))}::e^{-(\alpha_l,\phi(u'))}:=\nonumber\\
&&(iu-iu')^{\frac{(\alpha_k,\alpha_l)\beta^2}{2}}
(:e^{-(\alpha_k+\alpha_l,\phi(u))}:+\sum_{p=1}^{\infty}d^{kl}_p(u)(iu-iu')^p),
\nonumber
\end{eqnarray}
where $c^{kl}_p(u)$ and $d^{kl}_p(u)$ are operator-valued functions of $u$.   
Now one can see that only two types of terms 
can give the contribution of the order $\epsilon$ in 
$\mathbf{\bar{L}}^{(q)}(x_{m-1},x_{m})$ when $q\to 1$.
The first type consists of operators of the first order in $W_{\alpha_i}$ 
and the second type is formed
by the operators, quadratic in $W_{\alpha_i}$, which give contribution of the order
$\epsilon^{1\pm\beta^2}$ by virtue of operator product expansion.
Let's look on the terms of the second type in detail.\\
\hspace*{5mm}The terms of the second type appear from the quadratic products
of vertex operators arising from:\\ 
\hspace*{5mm}i)the composite roots (more precisely q-commutators of two fermionic roots),\\ 
\hspace*{5mm}ii)the quadratic terms of the q-exponentials which are present in the 
universal R-matrix.\\ 
\hspace*{5mm} At first we consider terms emerging from composite roots,
which have the following form (see Appendix):
\begin{eqnarray}\label{composite}
&&\frac{1}{a(\alpha_i+\alpha_j)(q-q^{-1})}[e_{\alpha_j},e_{\alpha_i}]_{q^{-1}}
(\int_{x_{m-1}}^{x_m}\d u_1
W_{\alpha_i}(u_1-i0)\nonumber\\ 
&&\int_{x_{m-1}}^{x_m}\d u_2 W_{\alpha_j}(u_2+i0)+
q^{b_{ij}}\int_0^{2\pi}\d u_2 W_{\alpha_j}(u_2-i0)
\int_0^{2\pi}\d u_1 W_{\alpha_i}(u_1+i0).
\end{eqnarray}
Using the fact that $q=e^{i\pi\frac{\beta^2}{2}}$ and that in the limit 
$\beta^2\to 0$, $a(\alpha_i+\alpha_j)\to -b_{ij}$, one can rewrite this 
as follows (leaving only terms that can give contribution to the first order in $\epsilon$:
\begin{eqnarray}
&&(2\pi i)^{-1}[e_{\alpha_j},e_{\alpha_i}]
\nonumber\\
&&\int_{x_{m-1}}^{x_m}\d u_1\int_{x_{m-1}}^{x_m}\d u_2\Big(\frac{1}{u_2-u_1+i0}-
\frac{1}{u_2-u_1-i0}\Big):e^{-(\alpha_i+\alpha_j,\phi(u_2))}:.
\end{eqnarray}
Now using the well known formula 
\begin{eqnarray}
\frac{1}{x+i0}-\frac{1}{x-i0}=-2i\pi\delta(x),
\end{eqnarray}
we obtain that (\ref{composite}) in the classical limit gives
\begin{eqnarray}
-[e_{\alpha_j},e_{\alpha_i}]
\int_{x_{m-1}}^{x_m}\d u:e^{-(\alpha_i+\alpha_j,\phi(u))}:.
\end{eqnarray}
Next let's consider the quadratic products arising from quadratic parts of q-exponentials of fermionic 
roots. They look as follows:
\begin{equation} 
\frac{-1}{(2)_{q^{-1}_{\alpha_i}}}\int_{x_{m-1}}^{x_m}\d u_1 
W_{\alpha_i}(u_1-i0)
\int_{x_{m-1}}^{x_m}\d u_2 W_{\alpha_i}(u_2+i0)e^2_{\alpha_i}. 
\end{equation}
One can rewrite this product via the ordered integrals:
\begin{equation} 
\frac{q^{b_{ii}}-1}{(2)_{q^{-1}_{\alpha_i}}}
\int_{x_{m-1}}^{x_m}\d u_1 W_{\alpha_i}(u_1)
\int_{x_{m-1}}^{u_1}\d u_2 W_{\alpha_i}(u_2)
e^2_{\alpha_i}.  
\end{equation}
In the limit $\beta^2\to 0$ we obtain (forgetting about the terms that 
could give contribution of the order $\epsilon^2$):
\begin{equation}  
-\frac{ib_{ii}\beta^2}{2}\int_{x_{m-1}}^{x_m}\d u_1\int_{x_{m-1}}^{u_1}\d u_2
(iu_1-iu_2)^{\frac{b_{ii}\beta^2}{2}-1}e^{-2(\alpha_i,\phi(u_2))}
e^2_{\alpha_i}.
\end{equation}
Therefore the final contribution is:
\begin{equation} 
-\int_{x_{m-1}}^{x_m}\d u e^{-2(\alpha_i,\phi(u))}
e^2_{\alpha_i}.
\end{equation}
Collecting now all the terms of order $\epsilon$ we find:
\begin{eqnarray} 
&&\mathbf{\bar{L}}^{(q)}(x_{m},x_{m-1})=1+\int_{x_{m-1}}^{x_m}\d u(
\sum_{f} W^F_{\alpha_f}(u)e_{\alpha_f}+
\sum_{b}W^B_{\alpha_b}(u)e_{\alpha_b}+\nonumber\\
&&\sum_{{f_1}\ge {f_2}}[e_{\alpha_{f_1}},e_{\alpha_{f_2}}]W^B_{\alpha_{f_1}+\alpha_{f_2}}(u))+
O(\epsilon^2).
\end{eqnarray}
Gathering the $\mathbf{\bar{L}}^{(q)}(x_{m},x_{m-1})$ it is easy to see that
in the $q\to 1$ limit\\ 
$\mathbf{\bar{L}}^{(q)}(x_{m},x_{m-1})$ is equal to:
\begin{eqnarray} 
&&\mathbf{\bar{L}}^{(cl)}(2\pi,0)=Pexp\int_0^{2\pi}\d u(\sum_{f} W^F_{\alpha_f}(u)e_{\alpha_f}+
\sum_{b}W^B_{\alpha_b}(u)e_{\alpha_b}-\nonumber\\
&&\sum_{f_1\ge f_2}[e_{\alpha_{f_1}},e_{\alpha_{f_2}}]W^B_{\alpha_{f_1}+
\alpha_{f_2}}(u)).
\end{eqnarray}
\hspace*{5mm}Defining then $\mathbf{L}^{(cl)}\equiv 
e^{\pi i p^i H^i}\mathbf{\bar{L}}^{(cl)}(2\pi,0)$ we find that it 
satisfies the quadratic Poisson bracket relation (\ref{rpoisson}) 
and defining the monodromy matrix 
$\mathbf{M}^{(cl)}\equiv e^{\pi i p^i H^i}\mathbf{L}^{(cl)}$ we again obtain 
the classical integrability condition (\ref{clintegr}).\\ 
\hspace*{5mm}Now let's find the L-operator which corresponds to the 
monodromy matrix defined above. Let's consider the following one:
\begin{equation}
\mathcal{L}_F=D_{u,\theta}-D_{u,\theta}\Phi^i (u,\theta)H^i-(\sum_{f}e_{\alpha_f}+
\sum_{b}\theta e_{\alpha_b}),  
\end{equation}
where $D_{u,\theta}=\partial_{\theta}+\theta\partial_u$ is a superderivative
and $\Phi^i$ are the classical superfields with the following Poisson 
brackets:
\begin{equation}
\{D_{u,\theta}\Phi^i(u,\theta), D_{u',\theta'}\Phi^j(u',\theta')\}=  
\delta^{ij}D_{u,\theta}(\delta(u-u')(\theta-\theta')).
\end{equation} 
Making a gauge transformation of the above L-operator one can arrive to the fields, satisfying 
classical version of super W-algebras with the commutation relations provided by the 
Poisson brackets \cite{gomes}, \cite{delduc}.\\ 
\hspace*{5mm}The associated ``fermionic'' linear problem can be reduced to the ``bosonic'' 
one. The linear problem
\begin{equation}
\mathcal{L}_F\Psi(u,\theta)=(D_{u,\theta}+N_1+\theta N_0)(\chi+\theta \eta),
\end{equation}
where $\Psi(u,\theta)=\chi+\theta \eta$, $N_1=\frac{i}{\sqrt{2}}\xi^i(u) H^i-\sum_{f}e_{\alpha_f}$, 
$N_0=-\partial_u\phi^i(u) H^i-\sum_b e_{\alpha_b}$,
can be reduced to the linear problem on $\chi$:
\begin{equation}
\mathcal{L}_B\chi(u)=(\partial_u+N_1^2+N_0)\chi(u).
\end{equation}
That is: 
\begin{equation}
\mathcal{L}_B=\partial_u-\partial_u\phi^i(u) H^i+(\frac{i}{\sqrt{2}}\xi^i(u) H^i
-\sum_{f}e_{\alpha_f})^2
-\sum_b e_{\alpha_b}.  
\end{equation}
One can easily see that the monodromy matrix for the corresponding linear problem is that 
described above.
\section{Integrals of Motion and Supersymmetry Invariance}
It is well known that (both classical and quantum) integrability conditions lead to the 
involutive family of (both local and nonlocal) integrals of motion (IM). For super- versions of 
these systems it is also known that sometimes it is possible to include supersymmetry generator 
\begin{eqnarray}
G_0=\beta^{-2}\sqrt{2}i^{-1/2}\int_0^{2\pi}du \phi'^l(u) \xi^l(u)
=\sum^r_{l=0}\sum_{n\in Z}\beta^{-2}\xi^l_n a^l_{-n}
\end{eqnarray}
in these series \cite{gervais}. Here we will show that the transfer matrix $\mathbf{t}^{(q)}(\lambda)
=str\mathbf{M}^{(q)}(\lambda)$ commute with $G_0$ if the simple root system is purely fermionic, 
that is:
\begin{eqnarray} 
\mathbf{t}^{(q)}(\lambda)=
str(\pi(\lambda)(e^{2i\pi P^k H^k}Pexp\int_0^{2\pi}\d u(\sum^r_{f=0} W^F_{\alpha_f}(u)e_{\alpha_f}))),
\end{eqnarray}
where $\pi$ denote some representation of the corresponding superalgebra in which the supertrace 
is taken.\\ 
\hspace*{5mm}We note the crucial property: 
\begin{eqnarray}\label{crucial}
[G_0,W^F_{\alpha}(u)]=-\partial_u W^B_{\alpha}(u)
\end{eqnarray}
Integrating over $u$ and multiplying by the appropriate coefficient one obtains:
\begin{eqnarray}
[G_0, e_{-\alpha_i}]=\frac{W^F_{\alpha_i}(0)-W^F_{\alpha_i}(2\pi)}{q-q^{-1}},
\end{eqnarray}
where $e_{-\alpha_i}$ is represented by the vertex operator 
(see previous Section). 
Next, we will use the important Proposition (Prop. 1, Sec. 3.1 of \cite{5}):
For the objects $A_i$, $B_i$, $\mathbf{I}$, satisfying the commutation relations
\begin{eqnarray} 
[\mathbf{I},e_{-\alpha_i}]=\frac{A_i-B_i}{q-q^{-1}}, \quad A_i e_{-\alpha_j}=q^{-b_{ij}}e_{-\alpha_j}A_i,
\quad B_i e_{-\alpha_j}=q^{b_{ij}}e_{-\alpha_j}B_i,
\end{eqnarray}
the following relation holds:
\begin{eqnarray}
[1\otimes \mathbf{I}, \bar{R}]=\bar {R}(\sum_i e_{\alpha_i}\otimes A_i)-
(\sum_i e_{\alpha_i}\otimes B_i)\bar{R}. 
\end{eqnarray}
Applying this relations to our case we find (identifying $A_i$ with 
$W^F_{\alpha_i}(0)$, $B_i$ with $W^F_{\alpha_i}(2\pi)$ and $\mathbf{I}$ with $G_0$):
\begin{eqnarray}\label{lemma}
[G_0,\mathbf{\bar{L}}^{(q)}(2\pi,0)]=\mathbf{\bar{L}}^{(q)}(2\pi,0)W^B(0)-W^B(2\pi)
\mathbf{\bar{L}}^{(q)}(2\pi,0),
\end{eqnarray} 
where 
\begin{eqnarray}
W^B(u)=\sum^{r}_{i=0}W^B_{\alpha_i}(u)e_{\alpha_i}.
\end{eqnarray}
Now using the property (\ref{commcircle}), periodicity properties of vertex operators:
\begin{eqnarray}
W^s_{\alpha}(u+2\pi)=q^{-(\alpha,\alpha)} e^{-2i\pi (\alpha, P)} W^s_{\alpha_i}(u+2\pi)
\end{eqnarray}
(here $s=B,F$) and cyclic property of the supertrace one obtains, multiplying both sides 
of (\ref{lemma}) by $e^{2i\pi P^k H^k}$ and taking the supertrace:
\begin{eqnarray} 
[G_0,\mathbf{t}^{(q)}]=0.
\end{eqnarray}
We note here that if there were bosonic simple roots in the construction of the transfer-matrix 
the above reasonings are no longer valid, because the corresponding bosonic vertex operators 
commuting with $G_0$ give the fermionic vertex operators associated with the same root vector 
(the same happens when we construct the superstring vertex operators \cite{GSW}), 
but not the total derivative as in (\ref{crucial}). It was already shown explicitly on the 
concrete examples that the hierarchies, based on the partly bosonic 
simple root systems are not invariant under the supersymmetry transformation (see e.g. 
\cite{penati}, \cite{evans}, \cite{gervais}).\\ 
\hspace*{5mm}The affine superalgebras which allow such root
systems are of the following type \cite{kac}: $A(m,m)^{(1)}=
sl(m+1,m+1)^{(1)}$, 
$A(2m,2m)^{(4)}=sl(2m+1,2m+1)^{(4)}$, 
$A(2m+1,2m+1)^{(2)}=sl(2m+2,2m+2)^{(2)}$, 
$A(2m+1,2m)^{(2)}=sl(2m+2,2m+1)^{(2)}$, 
$B(m,m)^{(1)}=osp(2m+1,2m)^{(1)}$, 
$D(m+1,m)^{(1)}=osp(2m+2,2m)^{(1)}$, 
$D(m,m)^{(2)}=osp(2m,2m)^{(2)}$, $D(2,1,\alpha)^{(1)}$.\\     
\hspace*{5mm} The involutive family of the (both classical and quantum) IM in the 
Toda field theories have the property, that the commutators of IM with 
the corresponding vertex operators reduce to the total derivatives 
\cite{feigin}: 
\begin{eqnarray}
[I_{l}, W_{\alpha_k}(u)]=\partial_u(:O^{(l)}_{\alpha_k}(u) 
W_{\alpha_k}(u):)=\partial_u\Theta^{(l)}_{k}(u),
\end{eqnarray}
where $O^{(k)}_{\pm}(u)$ is the polynomial of $\partial_u\phi^i(u)$, $\xi^i(u)$ and their derivatives.
In \cite{5} it was shown in the bosonic case 
(the proof is similar to the arguments above) that $I_{l}$ commute with the transfer matrix, the 
generalization of these arguments to the super-case is straightforward.\\ 
\hspace*{5mm}In the next two sections we will give two examples of 
the KdV hierarchies related to affine superalgebra 
$B(0,1)^{(1)}\equiv osp(1|2)^{(1)}$ (super-KdV) and twisted affine 
superalgebra $D(1,1)^{(2)}\simeq C(2)^{(2)}\equiv sl(2|1)^{(2)}\simeq 
osp(2|2)^{(2)}$ (SUSY N=1 KdV).
\section{Example 1: super-KdV hierarchy}
The super-KdV model \cite{kuper}, \cite{sKdV} is based on the following L-operator:
\begin{equation} 
D_{u,\theta}+D_{u,\theta}\Phi(u,\theta)h_{\alpha_0}-(e_{\alpha_1}+\theta e_{\alpha_0}),
\end{equation}
where $h_{\alpha_0}$, $e_{\alpha_1}$, 
$e_{\alpha_0}$ are the Chevalley generators 
of the upper Borel subalgebra of 
the $osp(1|2)^{(1)}$ which are 
taken in the evaluation representation that is $h_{\alpha_0}=-h$, $e_{\alpha_1}=iv_+$, 
$e_{\alpha_0}=\lambda X_-$, where $X_{\pm}$, $v_{\pm}$ 
and $h$ are the generators of $osp(1|2)$ superalgebra with the following commutation relations: 
\begin{eqnarray}
[h,X_\pm]&=&\pm 2X_\pm ,\quad [h,v_\pm]=\pm v_\pm ,\quad [X_+,X_-]=h,\\
\lbrack v_\pm,v_\pm\rbrack&=&\pm 2 X_\pm, \quad  [v_{+},v_{-}]=-h, \quad
[X_\pm,v_\mp]= v_\pm, \quad [X_\pm,v_\pm]=0.\nonumber
\end{eqnarray}
Here $p (v_\pm)=1$, $p (X_\pm)=0$, $p (h)=0$.
The classical monodromy matrix is:
 \begin{eqnarray}
\mathbf{M}(\lambda)&=&e^{-2\pi iph_{\alpha_0}}
P\exp\int_0^{2\pi} \d u\Big(\frac{i}{\sqrt{2}}\xi(u)e^{-\phi(u)}e_{\alpha_1}
-e^{-2\phi(u)}e_{\alpha_1}^2 \\
&+& e^{2\phi(u)}e_{\alpha_0}\Big).\nonumber
\end{eqnarray}
The involutive family of the integrals of motion which can be extracted 
from this monodromy matrix (more precisely they arise as a coefficients in the expansion in 
$\lambda^{-1}$ series of the 
trace of the logarithm of $\mathbf{M}$-matrix) can be expressed via the following fields: 
\begin{eqnarray}\label{eq:Miura}
U(u)=-\phi''(u)-\phi'^2(u)-\frac{1}{2}\xi(u)\xi'(u),\quad
\alpha(u)=\xi'(u) + \xi(u)\phi'(u)
\end{eqnarray}
generating the classical limit of the superconformal algebra under the Poisson brackets:
\begin{eqnarray}
\{U(u),U(v)\}&=&
 \delta'''(u-v)+2U'(u)\delta(u-v)+4U(u)\delta'(u-v),\\
\{U(u),\alpha(v)\}&=&
 3\alpha(u)\delta'(u-v) + \alpha'(u)\delta(u-v),\nonumber\\
\{\alpha(u),\alpha(v)\}&=&
 2\delta''(u-v)+2U(u)\delta(u-v).\nonumber
\end{eqnarray}
The integrals of motion are:
\begin{eqnarray}
&&I^{(cl)}_1=\int U(u)\d u,\\
&&I^{(cl)}_3=\int\Big(U^2(u)/2+\alpha(u)\alpha'(u)\Big)\d u,\nonumber\\
&&I^{(cl)}_5=\int\Big((U')^2(u)-2U^3(u)+8\alpha'(u)\alpha''(u)+12
\alpha'(u)\alpha(u)U(u)\Big)\d u,\nonumber\\
&&  .\qquad.\qquad.\nonumber
\end{eqnarray}
The second one $I^{(cl)}_3$ gives the super-KdV equation:
\begin{eqnarray}\label{s-KdV}
U_t=-U_{uuu}-6UU_u - 6\alpha\alpha_{uu},\quad
\alpha_t=-4\alpha_{uuu}-6U\alpha_u -3U_u\alpha.
\end{eqnarray}
However, the supersymmetry operator $G_0=\int_0^{2\pi}\d u\alpha(u)$ can not 
be included in the pairwise commuting 
IM that can be easily seen from the 
second equation of (\ref{s-KdV}) (i.e. $I^{(cl)}_3$ does not commute with $G_0$).   
Moving to the quantum case we find that the quantum analogue of the monodromy matrix is:
\begin{eqnarray}
\mathbf{M^{(q)}}(\lambda)=e^{-2\pi iPh_{\alpha_0}}
P\exp\int_0^{2\pi} \d u\Big(\frac{i}{\sqrt{2}}\xi(u):e^{-\phi(u)}:e_{\alpha_1}
+ :e^{2\phi(u)}:e_{\alpha_0}\Big),
\end{eqnarray}
where $h_{\alpha_0}$, $e_{\alpha_0}$, $e_{\alpha_1}$ 
are now the Chevalley generators of the $osp_q(1|2)^{(1)}$.
We did not put the letter $q$ over the P-exponential because this is the case 
when for values of $\beta^2$ from the interval (0,2) one can write the above 
object as a ``real'' P-exponential (represented via ordered integrals).     
Due to the presence of bosonic root $\e_{\alpha_0}$ the trace of the 
quantum monodromy matrix 
is not invariant under the supersymmetry transformation as it was in the classical case. 
\section{Example 2: SUSY N=1 KdV hierarchy}
The L-operator corresponding to the SUSY N=1 KdV model \cite{mathieu}, \cite{inami} 
is the following one:
\begin{eqnarray}
\mathcal{L}_F=D_{u,\theta} 
-D_{u,\theta}\Phi(u,\theta) h_{\alpha_1}-(e_{\alpha_0}+ e_{\alpha_1}),
\end{eqnarray}
where $h_{\alpha_1}$, $e_{\alpha_0}$, $e_{\alpha_1}$ are the Chevalley 
generators of the twisted affine Lie superalgebra $C(2)^{(2)}$ with such set of 
commutation relations:
\begin{eqnarray}
&&[h_{\alpha_1},h_{\alpha_0}]=0,\quad [h_{\alpha_0},e_{\pm\alpha_1}]= 
\mp e_{\pm\alpha_1},\quad [h_{\alpha_1},e_{\pm\alpha_0}]= 
\mp e_{\pm\alpha_0},\\
&&[h_{\alpha_i},e_{\pm\alpha_i}]=\pm e_{\pm\alpha_i},\quad
[e_{\pm\alpha_i}, e_{\mp\alpha_j}]=\delta_{i,j}h_{\alpha_i},\quad (i,j=0,1),
\nonumber\\
&&ad^{3}_{e_{\pm\alpha_0}} e_{\pm\alpha_1}=0, \quad
ad^{3}_{e_{\pm\alpha_1}} e_{\pm\alpha_0}=0\nonumber
\end{eqnarray}
Here $p(h_{\alpha_{0,1}})=0$, $p(e_{\pm \alpha_{0,1}})=1$, i.e. 
both simple roots are fermionic. The classical monodromy matrix is:
\begin{eqnarray}
\mathbf{M}&=&e^{2\pi i ph_{\alpha_1}}
P\exp\int_0^{2\pi} \d u\Big(\frac{i}{\sqrt{2}}
\xi(u)e^{-\phi(u)}e_{\alpha_1}\\
&-&\frac{i}{\sqrt{2}}
\xi(u)e^{\phi(u)}e_{\alpha_0}
-e^2_{\alpha_1}e^{-2\phi(u)}-
e^2_{\alpha_0}e^{2\phi(u)}-[e_{\alpha_1},e_{\alpha_0}]
\Big).\nonumber
\end{eqnarray}
The series of the integrals of motion starts with the following ones:
\begin{eqnarray}\label{IMSUSY}
I^{(cl)}_1&=&\frac{1}{2\pi}\int U(u)\d u,\\
I^{(cl)}_3&=&\frac{1}{2\pi}\int
\Big(U^2(u)+\alpha(u)\alpha'(u)/2\Big)\d u,\nonumber\\
I^{(cl)}_5&=&\frac{1}{2\pi}\int
\Big(U^3(u)-(U')^2(u)/2-\alpha'(u)\alpha''(u)/4-
\alpha'(u)\alpha(u)U(u)\Big)\d u,\nonumber\\
& &   .\qquad.\qquad.\nonumber
\end{eqnarray}
The fields $U$ and $\alpha$ are defined in terms of the free fields as in previous section, 
but now one can unify them into one superfield: 
\begin{equation}
\mathcal{U}(u,\theta)\equiv
D_{u,\theta}\Phi(u,\theta)\partial_u\Phi(u,\theta)-D_{u,\theta}^3
\Phi(u,\theta)=-\theta U(u)-i\alpha(u)/\sqrt{2}.
\end{equation}
The second IM from (\ref{IMSUSY}) generates the first nontrivial evolution 
equation, the SUSY $N$=1 KdV:
\begin{equation}
\mathcal{U}_t=-\mathcal{U}_{uuu}+3(\mathcal{U} D_{u,\theta}\mathcal{U})_u,  
\end{equation}
or in components:
\begin{eqnarray} 
U_t=-U_{uuu}-6UU_u - \frac{3}{2}\alpha\alpha_{uu},\quad \alpha_t=-4\alpha_{uuu}-3(U\alpha)_u.
\end{eqnarray}
Now one can see that unlike the previous example the supersymmetry generator 
$\int_0^{2\pi}\d u\alpha(u)$ commutes with $I^{(cl)}_3$ and can be included in the 
involutive family of IM \cite{gervais}. Moreover, the results of the Section 2 yield 
that the quantum monodromy matrix have the following form:
\begin{equation}\label{susymon}
\mathbf{M}=e^{2\pi iPh_{\alpha_1}}Pexp^{(q)}\int^{2\pi}_{0}\d u
(W_{-}(u)e_{\alpha_1} +W_{+}(u)e_{\alpha_0}),
\end{equation}
where $W_{\pm}=\int\d\theta:e^{\pm\Phi(u,\theta)}:$. Due to the fact that both roots 
are fermionic the supersymmetry generator commutes with the transfer matrix.\\
\hspace*{5mm}This SUSY $N$=1 KdV model was studied from a point of view  of 
the Quantum Inverse Scattering Method in \cite{npb}. There were constructed the analogues 
of Baxter's Q-operator, providing the following functional relations with the 
transfer-matrices:
\begin{eqnarray} 
\mathbf{t}_{\frac{1}{4}}(\lambda)\mathbf{Q}_{\pm}(\lambda)&=&\pm \mathbf{Q}_{\pm}
(q^{\frac{1}{2}}\lambda)\mp\mathbf{Q}_{\pm}(q^{-\frac{1}{2}}\lambda),\nonumber\\
\mathbf{t}_{\frac{1}{2}}(q^{\frac{1}{4}}\lambda)\mathbf{Q}_{\pm}(\lambda)&=&\mp\mathbf{t}_{\frac{1}{4}}
(q^{\frac{1}{2}}\lambda)\mathbf{Q}_{\pm}(q^{-\frac{1}{2}}\lambda)+\mathbf{Q}_{\pm}(q\lambda),
\end{eqnarray}
and the fusion relations between transfer-matrices in different representations:
\begin{eqnarray} 
\mathbf{t}_{j}(q^{\frac{1}{4}}\lambda)\mathbf{t}_{j}(q^{-\frac{1}{4}}\lambda)=
\mathbf{t}_{j+\frac{1}{4}}(\lambda)\mathbf{t}_{j-\frac{1}{4}}(\lambda)+(-1)^{4j},
\end{eqnarray}
which for the case when $q$ is a root of unity can be transformed into
the Thermodynamic Bethe Ansatz equations of $D_{2N}$ type \cite{zamolod}.\\   
\hspace*{5mm}It should be noted also that the associated 
Toda field theory is a well known $N$=1 SUSY sinh-Gordon model \cite{inami}, \cite{sin-g} with the action: 
\begin{equation}
\frac{1}{\beta^2}\int\d^2 u\d^2\theta (D_{u,\theta}\Phi\bar{D}_{\bar{u},\bar{\theta}}\Phi+m^2cosh(\Phi))
\end{equation}
\section*{Acknowledgements}
We are grateful to Professors F.A. Smirnov and M.A. Semenov-Tian-Shansky for useful discussions. 
The work was supported by the CRDF (Grant No. RUMI-2622-ST-04) and the Dynasty Foundation. 
\section{Appendix} 
The affine Lie superalgebra in has the following commutation relations between
its Chevalley generators ($H^i$ forms a basis in the Cartan subalgebra of 
the underlying simple Lie superalgebra and $e_{\pm\alpha_k}$ are the generators 
associated with positive and negative 
simple roots of the whole affine algebra) \cite{tolstkhor}:
\begin{eqnarray}\label{supercommrel}
&&[H^i, e_{\alpha_k}]=\alpha_k^i e_{\alpha_k}, \quad[e_{\alpha_k}, e_{-\alpha_l}]=\delta_{kl}
[h_{\alpha_k}]_q,
\quad ad_{e_{\pm\alpha_k}}^{1-a_{kj}}e_{\pm\alpha_j}=0,\\
&&[[e_{\pm\alpha_r},e_{\pm\alpha_s}]_q,[e_{\pm\alpha_r},e_{\pm\alpha_p}]_q]_q=0\nonumber 
\end{eqnarray}
if $(\alpha_r,\alpha_r)=(\alpha_s,\alpha_p)=(\alpha_r,\alpha_s+\alpha_p)=0$, in this case it is 
usually said that $\alpha_r$ is a ``grey'' root, which is between two roots 
$\alpha_{s}$, $\alpha_{p}$ on the Dynkin diagram \cite{scheunert}, \cite{leites}.
The definition of the super q-commutator is:
\begin{eqnarray}
ad^{(q)}_{e_{\alpha}}e_{\beta}=e_{\alpha} e_{\beta}-(-1)^{p(\alpha)p(\beta)}
q^{(\alpha,\beta)}e_{\beta}e_{\alpha},
\end{eqnarray}
where $p(\alpha)$ is equal to 1 when $\alpha$ is a fermionic root, and to 0 if $\alpha$ is a bosonic root.
The universal R-matrix for the contragradient 
Lie superalgebra of finite growth (affine algebra as a particular case) has the following structure:
\begin{equation}\label{R-matrix}
\mathbf{R}=K\bar{R}=K(\prod^{\to}_{\alpha\in{\Delta_{+}}}R_{\alpha}),
\end{equation}
where $\bar{R}$ is a reduced R-matrix and $R_{\alpha}$ are defined by the formulae:
\begin{equation}
R_{\alpha}=exp_{q_{\alpha}^{-1}}((-1)^{p(\alpha)}(q-q^{-1})(a(\alpha))^{-1}(e_{\alpha}\otimes e_{-\alpha})
\end{equation}
for real roots and 
\begin{equation}
R_{n\delta}=exp((-1)^{p(n\delta)}(q-q^{-1})(\sum^{mult}_{i,j}c_{ij}(n)
e^{(i)}_{n\delta}\otimes e^{(j)}_{-n\delta}))
\end{equation}
for pure imaginary roots. 
Here $\Delta_{+}$ is the reduced positive root system (the bosonic roots which are two 
times fermionic roots are excluded). 
The generators corresponding to the 
composite roots are defined according to the construction of the Cartan-Weyl basis given in \cite{tolstkhor}. 
For example the generators of the type $e_{\pm\alpha_{f_1}\pm\alpha_{f_2}}$
are constructed by means of the following q-commutators:
\begin{equation}
e_{\alpha_{f_1}+\alpha_{f_2}}=[e_{\alpha_{f_2}}, e_{\alpha_{f_1}}]_{q^{-1}},\quad 
e_{-\alpha_{f_1}-\alpha_{f_2}}=[e_{-\alpha_{f_1}}, e_{-\alpha_{f_2}}]_{q}.
\end{equation}  
The $a(\alpha)$ coefficients are defined as follows:
\begin{equation}
[e_{\gamma},e_{-\gamma}]=a(\gamma)\frac{k_{\gamma}-k^{-1}_{\gamma}}{q-q^{-1}}. 
\end{equation}
We will need the values of $a(\gamma)$ when $\gamma$ is equal to $\alpha_{f_1}+\alpha_{f_2}$, 
where $\alpha_{f_1}$ and $\alpha_{f_2}$ are fermionic simple roots:
\begin{equation}
a(\alpha_{f_1}+\alpha_{f_2})=\frac{q^{-b_{f_1 f_2}}-q^{b_{f_1 f_2}}}{q-q^{-1}}.
\end{equation}
The q-exponentials in (\ref{R-matrix}) are defined in the usual way:
\begin{eqnarray}
exp_q(x)&=&1+x+\frac{x^2}{(2)_q!}+...+\frac{x^n}{(n)_q!}+...=\sum_{n\ge0}\frac{x^n}{(n)_q!}\\
(a)_q&\equiv&\frac{q^{a}-1}{q-1},\quad q_{\alpha}\equiv(-1)^{p(\alpha)}q^{(\alpha,\alpha)}.\nonumber
\end{eqnarray}

\end{document}